# A Massive Young Star-Forming Complex Study in Infrared and X-ray: Mid-Infrared Observations and Catalogs

Michael A. Kuhn[1], Matthew S. Povich[1,2], Kevin L. Luhman[1,3], Konstantin V. Getman[1], Heather S. Busk[1], Eric D. Feigelson[1,3]

## ABSTRACT

*Spitzer* IRAC observations and stellar photometric catalogs are presented for the Massive Young Star-Forming Complex Study in the Infrared and X-ray (MYStIX). MYStIX is a multiwavelength census of young stellar members of twenty nearby ($d < 4$ kpc), Galactic, star-forming regions that contain at least one O star. All regions have data available from the *Spitzer* Space Telescope, consisting of GLIMPSE or other published catalogs for eleven regions and results of our own photometric analysis of archival data for the remaining nine regions. This paper seeks to construct deep and reliable catalogs of sources from the Spitzer images. Mid-infrared study of these regions faces challenges of crowding and high nebulosity. Our new catalogs typically contain fainter sources than existing *Spitzer* studies, which improves the match rate to *Chandra* X-ray sources that are likely to be young stars, but increases the possibility of spurious point-source detections, especially peaks in the nebulosity. IRAC color-color diagrams help distinguish spurious detections of nebular PAH emission from the infrared excess associated with dusty disks around young stars. The distributions of sources on the mid-infrared color-magnitude and color-color diagrams reflect differences between MYStIX regions, including astrophysical effects such as stellar ages and disk evolution.

*Subject headings:* methods: data analysis - stars: pre-main-sequence - infrared: stars - planetary systems: protoplanetary disks

[1]Department of Astronomy & Astrophysics, Pennsylvania State University, 525 Davey Lab, University Park, PA 16802

[2]Department of Physics and Astronomy, California State Polytechnic University, Pomona, California, 91768

[3]Center for Exoplanets and Habitable Worlds, Pennsylvania State University, University Park PA 16802

## 1. Introduction

A significant fraction of star formation activity in the Galaxy occurs in massive star-forming complexes, dominated by OB stars and containing thousands of young stars. Studies of the cluster mass function indicate that stars are more likely to be born in rich clusters than in small groups (e.g. Lada & Lada 2003; Fall et al. 2009; Chandar et al. 2011). Because of the importance of such clusters, the Massive Young Star-Forming Complex Study in the Infrared and X-ray (MYStIX) constructs a census of stars in twenty of the nearest ($d < 4$ kpc), Galactic massive star-forming regions (Feigelson et al. 2013) with the *Chandra X-ray Observatory*, the *Spitzer Space Telescope*, and ground based near-infrared (NIR) observatories. Studies in the IR and X-ray provide complementary pictures of populations of new stars in star-forming regions. The infrared (IR) images identify stars with circumstellar disks or infalling envelopes through infrared excess (IRE), but cannot distinguish disk-free cluster members from field stars. Meanwhile, X-ray images can detect both disk-bearing and disk-free stars, although the sensitivity to the former is lower (Getman et al. 2009; Stelzer et al. 2011). Thus, the combination of both X-ray and IR observations provide more complete and less biased samples of complex members than either waveband alone.

In this paper, we describe the observations and source catalogs used by the MYStIX project from the Infrared Array Camera (IRAC; Fazio et al. 2004) onboard the *Spitzer Space Telescope* (Werner et al. 2004). This instrument has four bands centered at 3.6, 4.5, 5.8, and 8.0 $\mu$m, which are useful for identifying IRE stars (e.g. Allen et al. 2004; Hartmann et al. 2005; Robitaille et al. 2006; Gutermuth et al. 2009). We use source catalogs produced by the pipeline of the Galactic Legacy Infrared Mid-Plane Survey Extraordinaire (GLIMPSE; Benjamin et al. 2003) for nine regions, and we measure new aperture photometry from archival IRAC data for nine regions. We adopt pre-existing IRE star catalogs for two additional regions, the Orion Nebula (Megeath et al. 2012) and the Carina Nebula (Povich et al. 2011). In addition to the data from *Spitzer*, the MYStIX project includes NIR data obtained by United Kingdom Infra-Red Telescope (King et al. 2013) and the Two Micron All Sky Survey (2MASS; Skrutskie et al. 2006) and X-ray data obtained by the *Chandra* X-ray Observatory (Kuhn et al. 2013a). The procedures for constructing the combined sample of X-ray selected member, IRE selected members, and spectrally selected OB members is described by Naylor et al. (2013), Povich et al. (2013), and Broos et al. (2013). We describe the available GLIMPSE data (Section 2) and our procedures for analyzing other archival images from IRAC (Section 3). We then discuss the distributions of probable members and field stars on color-magnitude and color-color diagrams (Section 4) and summarize our results (Section 5).

## 2. GLIMPSE Data

The GLIMPSE survey is a Legacy Science Program of NASA's *Spitzer Space Telescope* to study star formation in the disk of the Milky Way Galaxy (Benjamin et al. 2003; Churchwell et al. 2009). It contains six MYStIX regions – the Lagoon Nebula, the Trifid Nebula, NGC 6334, the Eagle Nebula, M 17, and NGC 6357 – within the 2°-wide strip along the Galactic equator (GLIMPSE I and II data releases). Furthermore, *Spitzer* images and photometry for RCW 38 and NGC 3576 come from the Vela-Carina survey (Majewski et al. 2007), using a similar observing strategy with mosaicking and photometric analysis performed with GLIMPSE pipeline.

For the GLIMPSE observations, every position was visited at least twice with 1.2 s integrations. The data-reduction pipeline produces image mosaics (v3.0) and point-source lists (v2.0) for all four IRAC bands, which are publicly available[1]. Photometry is obtained through point response function (PRF) fitting. A $5\sigma$ detection limit is used, corresponding to fluxes 0.2, 0.2, 0.4, and 0.4 mJy in the 3.6, 4.5, 5.8, and 8.0 $\mu$m bands, respectively (Benjamin et al. 2003). However, the detection sensitivity is lower in nebulous regions, and bright sources or regions with high backgrounds may be saturated. The GLIMPSE Catalog contains the sources with reliability $\geq$99.5%, and the GLIMPSE Archive contains all sources $\geq$5 $\sigma$ above the background level.

The GLIMPSE pipeline was run on a deep, high-dynamic range observation of the W 3 star-forming region (AOR 5050624). This GLIMPSE Catalog (Archive) contains >10,000 (>16,000 sources) shown in Table 2. These data were also reduced using the aperture photometry method ($\S$3) to compare results from the two methodologies.

## 3. New IRAC Analysis

The MYStIX project uses a combination of IRAC data from multiple provenances, as available. In addition to the GLIMPSE data described above, new analysis is performed on archival IRAC data for remaining MYStIX targets – the new catalogs have photometry extracted using aperture photometry (hereafter the aperture photometry catalogs) in contrast to GLIMPSE which makes use of PRF-fitting photometry. To guarantee that the MYStIX project has uniform data quality, our analysis includes a method comparison to study the effect of any biases produced by the variation in photometric method.

---

[1] http://www.astro.wisc.edu/sirtf/

The aperture photometry catalogs contain fainter sources than the GLIMPSE catalogs – which is primarily an effect of the longer observations from which the aperture photometry catalogs are derived, rather than an effect of differences in extraction method. This will improve the match rate to *Chandra* sources that are likely to be young stars, but a greater source density in the MIR catalogs will also increase the chance of incorrect matches (e.g. Naylor et al. 2013). Many such sources can be removed from further studies due to incongruous NIR/MIR photometry (Povich et al. 2013). In addition more extragalactic MIR sources are detected in the deeper catalogs, many of which have extragalactic X-ray counterparts. This is desirable because MIR properties (e.g. [4.5] > 13 mag) may help classify an X-ray source as being extragalactic (Harvey et al. 2007; Broos et al. 2013). Extragalactic IR sources without X-ray matches can be filtered out using cuts on the IR color-color diagram (Povich et al. 2013).

### 3.1. Observations

We obtained publicly available raw IRAC images from the *Spitzer* Heritage Archive[2] for nine MYStIX regions without GLIMPSE coverage. The target list and details of the Astronomical Observation Requests (AORs) are provided in Table 1. The IRAC field of view is $5\overset{.}{'}2 \times 5\overset{.}{'}2$, and various mapping and/or dither strategies were used for the IRAC observation programs included in this analysis. The camera spatial resolutions are FWHM $= 1\overset{.}{''}6$ to $1\overset{.}{''}9$ from 3.6 to 8.0 $\mu$m. Each field was observed in high dynamic range (HDR) mode where both 0.4 s and 10.4 s exposures are collected to provide unsaturated photometry for both brighter and fainter sources. Observations from different epochs are combined in our analysis.

We also analyzed archival data for M 17 and W 3 for comparison to the GLIMPSE data. For M 17 we analyzed images that are deeper than the *Spitzer* images from the GLIMPSE survey – useful for comparing relative sensitivities for the different catalogs. However, for W 3 we analyzed the same deep, archival *Spitzer* data using both the GLIMPSE and aperture photometry methods – useful for investigating biases of different data reduction methodologies.

---

[2]http://sha.ipac.caltech.edu/applications/Spitzer/SHA/

## 3.2. Mosaics

The basic calibrated data (BCD) products were created by the *Spitzer* pipeline. Image reduction and mosaicking was performed via WCSmosaic IDL package (Gutermuth et al. 2008). This procedure uses algorithms developed by the IRAC instrument team to mitigate image artifacts, such as jailbar, pulldown, muxbleed, and banding (Hora et al. 2004; Pipher et al. 2004). Long and short frames were merged to create an HDR mosaic for each target, and corrections are applied including cosmic ray identification, distortion corrections in each frame, derotation and subpixel offsetting, and background matching. Sub-pixel sampling was performed using the dithered images. The pixel size of the reduced mosaics is $0.''86 \times 0.''86$, which is $1/\sqrt{2}$ the native pixel width.

Mosaicked images of two sample targets – W 40 and NGC 2264 – are shown in Figure 1 in 3.6 and 8.0 $\mu$m bands. These examples demonstrate some of the variety of MYStIX regions in the MIR. For example, the stars in W 40 are centrally concentrated in the region where infrared nebulosity is highest, while the stars in NGC 2264 are divided into a number of subclusters and lie in regions with both high and low nebulosity (Feigelson et al. 2013, Kuhn et al. in preparation). Both regions have large amounts of absorption from dust – the dust absorption for W 40 is highest in a dust lane crossing the middle of the hour-glass structure (partially visible in the mosaics as infrared dark clouds), while the most highly absorbed stars in NGC 2264 are in subclusters that are embedded in their natal molecular cloud. The surface density of field stars also varies from region to region depending on the Galactic coordinates – W 40, $(l, b) = (28.8, +03.5)$, has a particularly high surface density, while the Flame Nebula, $(l, b) = (206.5, -16.4)$, has a much lower density. Several MYStIX regions, like NGC 2362, have almost no nebulosity around the star clusters because most of the molecular material has been removed.

## 3.3. Point-Source Photometry

The data reduction makes use of photometric procedures from Luhman et al. (2008a, 2008b, 2010), software from the Image Reduction and Analysis Facility (IRAF), codes from the IDL Astronomy Users Library (Landsman 1993), and visualization software from Broos et al. (2010). The methods are modified for the MYStIX regions, which are more distant, and have higher stellar crowding and nebulosity than the Taurus and Chamaeleon star forming regions treated by Luhman and colleagues. These modified methods have also been used by Getman et al. (2012) in their study of the IC 1396A star-forming region.

Source detection was performed on mosaicked images using the IRAF task STARFIND.

Some spurious detections appear in these initial lists, including statistically insignificant sources, IRAC image artifacts, the point-spread-function (PSF) wings of bright sources, and extended sources. The extended sources include peaks in the nebulosity, which are particularly prevalent in regions with bright, complex nebulosity, particularly in the 5.8 and 8.0 $\mu$m bands. Several strategies are used later to filter out unreliable sources. However, sources with bad or saturated pixels and duplicate detections are removed immediately.

Aperture photometry was performed on mosaicked images using the IRAF task PHOT. The targets lie near the Galactic plane and are crowded by field stars, so photometry is calculated for several small aperture sizes: 2-pixel ($1.7''$), 3-pixel ($2.6''$), 4-pixel ($3.5''$), and 14-pixel ($12.1''$) radii with an adjoining background annuli 1 pixel ($0.86''$) in width. The aperture/background sizes were chosen in accordance with the strategy of Lada et al. (2006), Luhman et al. (2008), Getman et al. (2009), and Getman et al. (2012); the latter finding no evident improvement in photometry using a "standard" 4-pixel-wide background instead of a 1-pixel-wide background used here. The zero-point IRAC magnitudes for the 14-pixel aperture are from Reach et al. (2005): ZP = 19.670, 18.921, 16.855, and 17.394 in the 3.6, 4.5, 5.8, and 8.0 $\mu$m bands, where $M = -2.5 \log(\mathrm{DN/sec}) + \mathrm{ZP}$. Aperture corrections for the other apertures are $0.640 \pm 0.016$, $0.725 \pm 0.012$, $0.968 \pm 0.030$, and $0.955 \pm 0.040$ for the 2 pixel aperture; $0.384 \pm 0.011$, $0.298 \pm 0.010$, $0.474 \pm 0.033$, $0.699 \pm 0.031$, for the 3 pixel aperture; $0.175 \pm 0.011$, $0.169 \pm 0.010$, $0.144 \pm 0.021$, and $0.222 \pm 0.025$ for the 4 pixel aperture in the 3.6, 4.5, 5.8, and 8.0 $\mu$m bands, respectively. Aperture sizes of 2, 3, or 4 pixels were assigned to each source depending on the crowding so that the error in flux is minimized. Calculation of aperture corrections, choice of apertures, and identification of crowded sources are discussed in Appendix A.

A cross-correlated IRAC catalog was generated from the four bands, using a threshold of $2''$ for matching (see Appendix B). To ensure the quality of the aperture photometry catalog in Table 3, only >5 $\sigma$ detections are reported and every object must be detected in both 3.6 and 4.5 $\mu$m bands to be included. Sources with high levels of contamination from a neighboring source (>100% of the source flux) are excluded. An archive of the less reliable sources detected at >3 $\sigma$ is also preserved (see Appendix C).

## 3.4. IRAC Source Lists

Table 3 presents the aperture photometry catalog for the nine MYStIX fields analyzed here. Columns in Table 3 include positions, IRAC band magnitudes and their uncertainties, and aperture size flags. The uncertainty incorporates the statistical uncertainty calculated by aperture photometry, added in quadrature to a ~0.02 mag uncertainty in the calibration of IRAC (Reach et al. 2005), and a ~0.01 mag uncertainty in the aperture correction. YSO variability of ~0.05 mag to ~0.2 mag may contribute to photometric scatter from one observation to another (Morales-Calderón et al. 2009). The aperture flag indicates which photometric aperture size is used and whether errors due to crowding exceed 10% of the flux in the 3.6 $\mu$m band.

Table 4 summarizes the aperture photometry and GLIMPSE catalogs for each region – the total number of sources is given in Column 8. Variation in number of sources is strongly related to the size of the field of view (Column 3), its Galactic coordinates (Column 2), and the depth of the observation. Of the aperture photometry sources, only 25% are detected in the 5.8 $\mu$m band and 18% are detected in the 8.0 $\mu$m band (14% are detected in both bands). The distribution of aperture sizes is: 13% use 4 pixels, 11% use 3 pixels, and 76% use 2 pixels (43% of catalog sources have flags indicating crowding).

### 3.4.1. Completeness Limits

Photometric reliability and completeness are spatially variable in regions with bright, structured background. Histograms of source flux provide a rough estimate of spatial completeness due to the sharp turnover beyond the completeness limit. The histograms of IRAC magnitudes for each field are shown in Figure 2 with bin widths of 0.2 mag. The completeness limits, estimated to be the center for the bin preceding the bin with the most sources, are listed in Table 4. However, these values do not hold where there is high nebulosity. For a typical field – a cluster age of 2 Myr, a distance of 2 kpc, and a completeness limit of $[3.6]_c = 16.0$ – disk-free members would be detectible down to ~0.1 $M_\mathcal{S}$ in regions with low nebulosity for the pre-main-sequence models of Siess et al. (1997). Completeness limits for IRE sources may occur at lower magnitudes due to selection effects when multiple bands and their errors are combined.

### 3.4.2. Photometric Quality in Comparison to GLIMPSE

It is helpful to compare the aperture photometry to the GLIMPSE photometry to examine how the various MIR challenges in MYStIX regions, such as nebulosity and crowding, affect the catalogs. For comparative purposes, aperture photometry catalogs were produced for W 3 using the same deep HDR data used by GLIMPSE, and for M 17 using a deeper observation than GLIMPSE (see Table 1).

The aperture photometry catalog for W 3 is sensitive to ~1 magnitude deeper than GLIMPSE. More than 90% of GLIMPSE W 3 sources are detected at $>5\sigma$ by our aperture photometry method, and many of the undetected sources are the dimmer components of close double sources.

Figure 3 shows difference in measurements of magnitude for the two W 3 catalogs. As expected, scatter increases with magnitude: the root mean square (RMS) 3.6 $\mu$m band residuals are 0.06 mag for bright sources ([3.6] < 12 mag) and 0.17 mag for dim sources ([3.6] > 12 mag). There is also a slight systematic shift in the 3.6 $\mu$m band of +0.01 mag for bright sources and +0.05 mag for dim sources. Overall, the RMS residuals are 0.17 mag, 0.18 mag, 0.21 mag, and 0.22 mag in the 3.6, 4.5, 5.8, and 8.0 $\mu$m bands, respectively. Furthermore, the scatter increases for smaller aperture sizes: in the 3.6 $\mu$m band for sources with [3.6] < 12 mag the RMS residuals are 0.05 mag for 4 pixels, 0.06 mag for 3 pixels, 0.10 mag for 2 pixels (without high crowding or 0.14 for 2 pixels with high crowding). This is consistent with the aperture comparisons in the IC 1396A field performed by Getman et al. (2012).

Similar trends are seen in for M 17 (Figure 4), for which we compare aperture photometry of HDR observation of M 17 to the shallower GLIMPSE survey data. For bright sources ([3.6] < 12 mag) the RMS residuals are 0.14 mag, 0.13 mag, 0.22 mag, and 0.29 mag, and for dim sources ([3.6] > 12 mag), these are 0.22 mag, 0.23 mag, 0.33 mag, and 0.50 mag in the 3.6, 4.5, 5.8, and 8.0 $\mu$m bands, respectively. For bright sources ([3.6] < 12 mag) in the two M 17 catalogs, 95% of the photometry is within 0.2 mag in the 3.6 $\mu$m band, which drops to 85% for the 8.0 $\mu$m band. For dim sources, the consistency is ~73% for the 3.6 $\mu$m band, dropping to ~68% for the 8.0 $\mu$m band. A comparison of Figures 3 and 4 shows significantly more scatter for M 17 than for W 3; this is due to the shallowness of the M 17 GLIMPSE observation, which demonstrates, not only the benefits of the deeper HDR observations, but that the effect of net exposure duration is more important than any bias due to different photometry extraction methods in MYStIX.

Figure 5 shows a band-by-band comparison of signal-to-noise from the two W 3 catalogs. Sources that are in one catalog but not in the other are also indicated. As expected, the aperture photometry catalog includes somewhat more sources with low signal-to-noise (5 < $S/N$ < 10) than in the GLIMPSE Catalog. For sources in both catalogs, the signal-to-noise values lie near $y = x$ (for the 3.6 and 4.5 $\mu$m bands, signal-to-noise from aperture photometry is on average 1.5 times smaller) but can vary by a factor of ~2.

In Figure 5, the overlap between the catalogs decreases with increasing wavelength bands, with most sources in common in the 3.6 $\mu$m band and fewest in the 8.0 $\mu$m band. This may be an effect of marginally detectable sources in the nebulous and crowded W 3 region; detection sensitivity decreases with longer wavelength bands due to lower efficiency of the detectors, less photospheric flux, and higher nebulosity. Both GLIMPSE catalogs and the aperture photometry catalogs may capture only a fraction of the sources near the sensitivity limit, but the sources that are detected are not necessarily the same sources.

Sources that are likely to be spurious detections of nebulosity (see §3.5) are also indicated. Many of these are have low signal-to-noise (particularly sources not detected by GLIMPSE); however, a few have high signal-to-noise values.

## 3.5.    Contamination by Nebulosity

The 3.6, 5.8, and 8.0 $\mu$m bands are tuned to emission bands of polycyclic aromatic hydrocarbons (PAH; Reach et al. 2006) excited by the ultraviolet light of OB stars in the MYStIX fields. This results in extremely bright nebulosity in these bands when observing massive star-forming complexes. This nebulosity is often similar in surface brightness to young stars at the resolution of *Spitzer*, making it difficult to distinguish between point sources and contaminants due to confusion. Ideally, peaks in the nebulosity should be filtered out by STARFIND using the source profile, but this often fails, and this judgement is often difficult to make by eye as well. The level of nebulosity ranges from almost none in NGC 2362 to levels at which IRAC point-source photometry is impossible. In Figure 1, the 8 $\mu$m nebular emission can be seen to be higher in W 40 than in NGC 2264. For the regions with most nebulosity, including W 40, RCW 36, and W 3, source detection sensitivity can be severely limited.

Nebular contamination may result in two types of spurious entries in the aperture photometry catalog: patches of nebulosity with emission in all four bands that mimic stars, and false matches between stellar sources in the 3.6 and 4.5 $\mu$m bands and nebular patches in the 5.8 and 8.0 $\mu$m bands. These effects are also present in GLIMPSE (Povich et al. 2013), but are more prevalent here due to both the improved identification of extended sources using GLIMPSE's PSF fitting and the higher sensitivity of our aperture extraction method. Selecting sources with high signal-to-noise (reported in Table 3) can produce lists of more reliable sources, since nebulous sources are likely to have background extraction with greater pixel-to-pixel variation. However, nebulous sources can occasionally have small measurement errors because these sources can be very bright.

Nebulosity at 5.8 or 8.0 $\mu$m can make a source appear red, but the colors are distinct from the colors of young stellar objects. The [4.5]-[5.8] vs. [5.8]-[8.0] diagram can be used to separate these sources from stellar sources (Povich et al. 2013). In Figure 6 this diagram is shown for NGC 2264 and W 40, with sources color-coded by signal-to-noise <10 (green) and $>10$ (black). This plot includes only sources with photometric data in all four bands, which is the minority of MYStIX MIR sources, and is strongly biased toward IRE sources or sources with nebulosity in the 5.8 and 8.0 $\mu$m bands. Stars without IRE are centered near (0,0) on the diagram, while stars with IRE are shifted slightly to the upper right. However, there is another population with $[4.5] - [8.0] \gtrsim 1.6$ and $[5.8] - [8.0] \gtrsim 0.5$ that are likely due to nebulosity (Povich et al. 2013). Nearly all of these sources have $S/N < 10\sigma$ in at least one band. However, a number of low signal-to-noise sources also have colors consistent with stellar photospheres or young stellar objects.

Figure 7 shows the 8.0 $\mu$m sources from the aperture photometry catalog and from the literature plotted on the 8.0 $\mu$m image for NGC 2264 (Sung et al. 2009), NGC 2362 (Currie et al. 2009), Rosette (Balog et al. 2007), and NGC 1893 (Caramazza et al. 2008). In this comparison there are examples of detections in the aperture photometry catalog that are not in theirs and vice versa. It is difficult to determine through visual inspection of the 8 $\mu$m images which of these are real. The X-ray sources from Kuhn et al. (2013a) are also plotted – most of which are young stars – and these show that many of the new 8 $\mu$m sources found from aperture photometry coincide with X-ray sources. This phenomenon is particularly strong for NGC 1893, which has a long X-ray exposure but is distant, so many of the young stars are near the detection threshold in the MIR. Thus, we gain many new MIR counterparts for cluster members by using these more sensitive catalogs.

## 4. Classes of MIR Sources

The MYStIX MIR catalogs include young stellar members of the star-forming complex (with and without IRE), non-member point sources (field stars, extragalactic sources, shock emission), and spurious sources. MYStIX IRE Sources (MIRES; Povich et al. 2013) are identified using the MIR and NIR photometry, and a list of MYStIX Probable Complex Members (MPCM; Broos et al. 2013) is generated using X-ray selected members, IRE selected members, and spectroscopic OB stars. The distributions of these classes of sources on the MIR color-magnitude and color-color diagrams can give insight into how the MYStIX census is affected by the MIR catalog properties – properties such as the completeness limits, uncertainties on photometry, and spurious sources. Furthermore, these diagrams reveal global differences from region to region in terms of member populations, disk evolution, and star-formation environments.

### 4.1. MIR Color-Magnitude Diagram

Figure 8 shows the [3.6] vs. [3.6] − [4.5] diagrams for the nine regions analyzed here. The distributions of disk-free MPCMs (green circles) and MIRES sources (red circles) are plotted along with unclassified MIR sources (grey points). The $A_K = 2$ reddening vector points to the lower right. Selection effects such as the size of the sample, the completeness limit, and larger photometric uncertainties for faint points can be seen for each region. The locus of disk-free members overlaps with the locus of field stars, and the 3.6 $\mu$m band magnitude relates to stellar mass. In older or more distant regions, the dereddened, disk-free isochrones are shifted downwards on the plot (e.g. Roccatagliata et al. 2011, their Figure 7). The MYStIX MIR catalogs are typically deeper than the MYStIX X-ray catalogs, so X-ray selected MPCMs have a lower 3.6 $\mu$m band magnitude completeness limit.

There is a population of stars that shows [3.6] − [4.5] excess on this diagram, many of which are identified as IRE sources in the MIRES catalog (red circles). However, the distribution of these sources varies from region to region. In some cases there are many sources with [3.6] −[4.5] > 1.5 (including NGC 2264, Rosette, and DR 21) while for other cases nearly all IRE sources have [3.6] − [4.5] < 1.0 (including NGC 2362, W 4, and NGC 1893). Other fields (like Flame, W 40, and RCW 36) are intermediate. NGC 2264, Rosette, and DR 21 all have young embedded clusters, while the clusters in NGC 2362, W 4, and NGC 1893 are mostly lightly absorbed. The sources with [3.6] − [4.5] > 1.5 are mostly clustered within the embedded clusters identified by (Kuhn et al. in preparation). The different distributions of [3.6] − [4.5] colors for different regions is primarily due to effects of IRE emission rather than reddening from dust. It would require ~100 mag of absorption in the $V$ band to cause a reddening of 1 mag in [3.6] − [4.5], and most of the cluster members in our sample do not have this much reddening (Povich et al. 2013; Broos et al. 2013). Therefore, the larger [3.6] −[4.5] excesses in some regions is likely to be an age effect of disk evolution.

### 4.2. MIR Color-Color Diagram

Figure 9 shows the [3.6]-[4.5] vs. [4.5]-[8.0] color-color diagram for sources in the Rosette Nebula. Here, the reddening vector is nearly vertical, and IRE from disks or envelopes appears as an excess in both colors. Sources contaminated by PAH nebulosity will have large [4.5]-[8.0] values but [3.6]-[4.5] colors near or below 0, so they can be distinguished from young stars. There are a variety of possible cuts on the color-color diagram that are designed to identify young stars, and the IRE selection polygon from Simon et al. (2007) is shown as an example that uses this color-color plot. But, the IRE sources found by Povich et al. (2013) in the MYStIX IR catalogs are a somewhat different set of sources than are found using the other schemes[3]. Figure 10 shows color-color diagrams for each region with points color-coded by results of the classification done by Povich et al. (2013) and Broos et al. (2013), which includes cluster members with and without IRE in addition to various types of contaminants and spurious sources.

Reddened stellar photosphere fitting may quickly remove a large fraction of IR catalog sources from a list of possible IRE stars and relies on well understood field-star photospheric models. The procedures for fitting these stars using photometry in seven NIR and MIR bands are described in Povich et al. (2013). Some of the sources rejected for being insignificantly different from the reddened photospheric model (shown in black in Figure 10) would have been selected as IRE stars by the color cuts from Simon et al. (2007).

The color-color polygon used by Simon et al. (2007), and other color-based decision trees, have both false positives and false negatives with respect to the more elaborate analysis of the infrared spectral energy distributions by Povich et al. (2013) for the MYStIX analysis. Many of the stars lying in the polygon do not satisfy the more conservative criteria for disk-bearing young stars adopted by Povich et al. These sources may often be nebular (rather than stellar) sources in the 8.0 $\mu$m band, as they are more common in the W 40 and DR 21 where the PAH contamination is high. The MPCM source lists also show a small population of disk-free stars (green circles in Figure 10) with MIR colors likely to arise from PAH nebulosity. These may be faulty matches between true X-ray sources and spurious MIR PAH sources.

## 5.  Summary

This work describes the *Spitzer* IRAC observations and source catalogs that will be used by the MYStIX project. These data include nine regions where archival data is available, and we perform aperture photometry on the HDR observations. The MYStIX MIR catalogs will be combined with X-ray (Kuhn et al. 2013, Townsley et al. in preparation) and NIR (King et al. 2013) for a multiwavelength study of star formation in massive young star-forming complexes (Feigelson et al. 2013; Naylor et al. 2013; Povich et al. 2013; Broos et al. 2013). The aperture photometry catalogs are typically deeper and have higher photometric precision than typical GLIMPSE fields or other available catalogs of the same regions.

---

[3]The IRE analysis in Povich et al. (2013) adapts SED fitting methods of Robitaille et al. (2007) and color-cuts of Gutermuth et al. (2009). The effects of these modifications on IRE selection are discussed there.

In addition, the MYStIX project makes use of the GLIMPSE photometry for ten regions (including a deep catalog for W 3 presented here). Furthermore, the MYStIX study of the Orion Nebula and Carina Nebula uses stellar membership censuses from the literature, and we do not reanalyze *Spitzer* data for these regions.

There are a total of ~750,000 infrared sources in the aperture photometry catalogs. Photometry is extracted using variable aperture size depending on source crowding. We use a $>5\sigma$ detection threshold, require sources to be detected in both the shorter wavelength IRAC bands, and clean the catalog of various instrumental and data-processing effects. In the study of MYStIX X-ray sources, lower reliability ($>3\sigma$) detections will also be included (the aperture-photometry archive) because the presence of an X-ray counterpart provides corroborating evidence for a source's legitimacy.

We investigate a variety of possible photometric problems empirically by comparing detection rates, fluxes, and flux uncertainties for the aperture photometry catalogs to other available catalogs. A particular problem we encounter is spurious sources due to nebulosity, which affect all bands, but particularly strongly affect the 5.8 and 8.0 $\mu$m bands. These sources are difficult to eliminate completely through signal-to-noise cuts, although they usually affect detections with $5 < S/N < 10$. However, color-color diagrams can be used to separate colors associated with PAH nebulosity from IRE candidate young stars. In addition, spatial completeness limits vary across the field due to strong variation in nebulosity and crowding.

Finally, we present MIR color-magnitude and color-color diagrams showing the locus of MYStIX Probable Cluster Members (Broos et al. 2013) in comparison to the field stars. The nine MYStIX regions studied here show considerable differences in the distribution of IRE stars in these plots.

The MYStIX project is supported at Penn State by NASA grant NNX09AC74G, NSF grant AST-0908038, and the *Chandra* ACIS Team contract SV4-74018 (G. Garmire & L. Townsley, Principal Investigators), issued by the *Chandra* X-ray Center, which is operated by the Smithsonian Astrophysical Observatory for and on behalf of NASA under contract NAS8-03060. We thank Marilyn R. Meade and Brian L. Babler for providing us with the reduced images and photometry for W 3. We thank the anonymous referee for closely reading the manuscript and providing useful comments and suggestions. This work is based on observations made with the Spitzer Space Telescope, obtained from the NASA/ IPAC Infrared Science Archive, both of which are operated by the Jet Propulsion Laboratory, California Institute of Technology under a contract with the National Aeronautics and Space Administration. This research has also made use of SAOImage DS9 software de-veloped by Smithsonian Astrophysical Observatory and NASA's Astrophysics Data System Bibliographic Services.

## A. Photometric Aperture Sizes

Larger aperture sizes are favored for stars that are not in crowded regions because they have less "aperture noise," which is an effect of resampling the pixelated image into the aperture (Shahbaz et al. 1994). This effect can lead to several percent error in flux measurement using our two-pixel apertures, which is independent of source flux. Thus, we use simulations of artificial sources to determine the largest aperture size that will not cause inaccurate flux measurements due to crowding.

Using the IRAC PSF[4], pairs of point sources were simulated with various separation angles (1 to 20 pixels), orientations, and flux differences ($10^{-3}$ to $10^3$), and their photometry was extracted using PHOT to investigate the effect of nearby neighbors on flux measurements. This was performed using the same 2-pixel, 3-pixel, and 4-pixel apertures with 1-pixel-wide background annuluses that were used for the photometric analysis. For each separation angle and difference in difference in flux, the largest aperture (2-pixels, 3-pixels, or 4-pixels) is chosen that keeps contamination from a nearby source <5% of the true flux; these choices are listed in Table 5 along with the associated error in flux. For sources in the *Spitzer* catalogs using the 2-pixel aperture, sources with flux errors larger than 10% are flagged and sources with flux errors larger than 100% of the true flux are excluded from the catalog. Getman et al. (2012, §2.2) has shown that the use of small aperture extraction produces negligible bias in derived magnitudes, but leads to reduced photometric precision. These larger photometric uncertainties are incorporated in the aperture photometry catalogs given in the present paper.

Aperture corrections for the 2-pixel, 3-pixel, and 4-pixel apertures are calculated for each field with respect to magnitudes derived for the 14 pixel aperture, which is assumed to contain all the light. Corrections are found by comparing magnitudes for the 14-pixel aperture to the 4-pixel aperture, the 4-pixel aperture to the 3-pixel aperture, and the 4-pixel aperture to the 2-pixel aperture. Typically, bright sources that are not saturated and do not show signs of anomalous magnitudes in either aperture are used, and the correction is the median difference in calculated magnitudes. For IRAC channels 3 and 4, there is an artifact in the PSF of bright sources, so dimmer sources are used for the calibration.

---

[4]http://irsa.ipac.caltech.edu/data/SPITZER/docs/irac/calibrationfiles/psfprf/

### B. Matching Sources from Different IRAC Bands

The matching algorithm is based on the Delaunay triangulation, which is a computationally efficient strategy for identifying neighboring points in a set of points (see review by de Berg et al. 2008). To match sources from two bands, the sets of source positions from the two bands are combined, the Delaunay triangulation is constructed for the union of both sets, and matches are obtained as a subset of edges in the triangulation that join points from both bands and are shorter than a threshold length. In ambiguous cases, preference is given to the smallest separation.

Here, we use a matching radius of $2^{ll}$, and match sources in the 4.5, 5.8, and 8.0 $\mu$m bands to the 3.6 $\mu$m band sources. An astrometric correction is applied to the entire field in the 4.5, 5.8, and 8.0 $\mu$m bands based on the median offsets relative to 3.6 $\mu$m band positions, and matching is performed again with the improved positions. Right ascensions and declinations are reported for the 3.6 $\mu$m source.

The PSF has a similar size for all bands, so it is uncommon for matches to be ambiguous within the IRAC aperture photometry catalogs. Ambiguous matches do commonly occur when comparing IRAC catalogs with UKIRT (King et al. 2013) or *Chandra* (Kuhn et al. 2013, Townsley et al. in preparation) catalogs later in the MYStIX analysis. For this reason, a probabilistic approach to cross-waveband source matching is developed by Naylor et al. (2013). However, occasional intra-IRAC matching errors do arise from matches between real 3.6 and 4.5 $\mu$m point sources to peaks in the nebulosity at 5.8 or 8.0 $\mu$m within the $2^{ll}$ radius. These are filtered out using a conservative analysis of the near- and mid-infrared spectral energy distributions later in the MYStIX analysis (Povich et al. 2013).

### C. Archive of Lower-Reliability IRAC Sources

For the matching of IR sources to X-ray sources, the MYStIX project makes use of some MIR counterparts that have less-reliable photometry, including measurements with $5 > S/N > 3$ and measurements that may be contaminated by bright, neighboring sources. The extra information provided by X-ray selection allows us to take this bifurcated approach to MIR reliability criteria. When we are considering the tens-of-thousands to hundreds-of-thousands of MIR sources without X-ray counterparts, rare photometric errors could produce numerous false IRE-star candidates, so the reliability criteria must be strict. In contrast, there are orders-of-magnitude fewer MIR sources with X-ray counterparts – which already have a good chance of being young stars by virtue of X-ray emission – so rare photometric errors will be less important. The same approach is taken with the GLIMPSE data with respect to the highly reliable GLIMPSE Catalog and the less reliable GLIMPSE Archive. MIR photometry for MYStIX sources that used data from the GLIMPSE Archive or aperture-photometry archive is provided by Broos et al. (2013, their Table 2).

Table 1.  IRAC Observing Log

| Target (1) | AOR (2) | PID (3) | Center (J2000.0) (4) | FOV (deg$^2$) (5) | Integration (s pix$^{-1}$) (6) | Date (UT) (7) |
|---|---|---|---|---|---|---|
| Flame | 8770816 | 43 | 05 41 57.3 −01 25 04 | 0.54 | 10.4 | 2004-10-27 |
| | 8771072 | 43 | 05 41 41.0 −01 51 31 | 0.11 | 10.4 | 2004-02-17 |
| | 8771328 | 43 | 05 41 41.0 −01 51 31 | 0.11 | 10.4 | 2004-10-27 |
| | 8771584 | 43 | 05 41 56.5 −01 25 00 | 0.53 | 10.4 | 2004-02-17 |
| | 8771840 | 43 | 05 42 06.7 −02 12 34 | 0.51 | 10.4 | 2004-02-17 |
| | 8772096 | 43 | 05 41 55.6 −01 25 21 | 0.54 | 10.4 | 2004-10-27 |
| | 8772352 | 43 | 05 41 41.5 −01 51 06 | 0.11 | 10.4 | 2004-02-17 |
| | 8772608 | 43 | 05 42 07.4 −02 12 29 | 0.51 | 10.4 | 2004-10-27 |
| | 8772864 | 43 | 05 41 41.5 −01 51 06 | 0.11 | 10.4 | 2004-10-27 |
| | 8773120 | 43 | 05 42 06.6 −02 12 08 | 0.50 | 10.4 | 2004-02-17 |
| | 8773376 | 43 | 05 41 56.4 −01 25 25 | 0.54 | 10.4 | 2004-10-27 |
| | 8773632 | 43 | 05 42 05.7 −02 12 13 | 0.51 | 10.4 | 2004-10-27 |
| W 40 | 19958016 | 30574 | 18 30 32.0 −01 33 50 | 0.55 | 20.8 | 2006-10-27 |
| | 19958272 | 30574 | 18 32 50.8 −01 33 48 | 0.55 | 20.8 | 2006-10-27 |
| | 19998464 | 30574 | 18 28 29.7 −02 37 06 | 0.56 | 20.8 | 2006-10-28 |
| | 19998720 | 30574 | 18 30 24.3 −02 37 03 | 0.56 | 20.8 | 2006-10-28 |
| | 19999488 | 30574 | 18 32 34.7 −02 37 00 | 0.56 | 20.8 | 2006-10-28 |
| | 20002560 | 30574 | 18 30 31.7 −01 33 46 | 0.56 | 20.8 | 2006-10-27 |
| | 20003072 | 30574 | 18 32 50.5 −01 33 44 | 0.56 | 20.8 | 2006-10-27 |
| | 20018688 | 30574 | 18 28 30.0 −02 37 10 | 0.56 | 20.8 | 2006-10-28 |
| | 20018944 | 30574 | 18 30 24.6 −02 37 07 | 0.56 | 20.8 | 2006-10-28 |
| | 20019200 | 30574 | 18 32 35.0 −02 37 04 | 0.56 | 20.8 | 2006-10-28 |
| RCW 36 | 15990016 | 20819 | 08 59 27.5 −43 45 27 | 0.03 | 93.6 | 2006-05-02 |
| NGC 2264 | 3956480 | 37 | 06 40 54.9 +09 37 08 | 0.50 | 10.4 | 2004-03-06 |
| | 3956736 | 37 | 06 40 54.9 +09 37 08 | 0.50 | 10.4 | 2004-10-08 |
| | 3956992 | 37 | 06 40 54.9 +09 37 08 | 0.50 | 10.4 | 2004-03-06 |
| | 3957248 | 37 | 06 40 54.9 +09 37 08 | 0.50 | 10.4 | 2004-10-08 |
| NGC 2362 | 20590592 | 30726 | 02 26 34.4 +62 00 43 | 0.07 | 124.8 | 2007-09-14 |
| DR 21 | 22498048 | 40184 | 20 37 58.2 +42 40 30 | 1.08 | 31.2 | 2007-11-22 |
| Rosette | 10878464 | 3394 | 06 34 27.00 +04 12 23 | 1.31 | 31.2 | 2005-03-26 |
| | 20591872 | 30726 | 06 31 58.0 +04 55 46 | 0.08 | 124.8 | 2006-10-30 |

Table 1—Continued

| Target (1) | AOR (2) | PID (3) | Center (J2000.0) (4) | FOV (deg$^2$) (5) | Integration (s pix$^{-1}$) (6) | Date (UT) (7) |
|---|---|---|---|---|---|---|
|  | 21922560 | 40359 | 06 32 18.0 +04 52 00 | 0.29 | 31.2 | 2009-04-21 |
|  | 21922816 | 40359 | 06 32 18.0 +04 52 00 | 0.33 | 31.2 | 2009-04-21 |
|  | 21923072 | 40359 | 06 32 18.0 +04 52 00 | 0.43 | 31.2 | 2009-04-21 |
|  | 21923328 | 40359 | 06 34 27.0 +04 12 23 | 0.48 | 31.2 | 2009-04-21 |
|  | 21924608 | 40359 | 06 33 43.1 +04 46 58 | 0.07 | 31.2 | 2007-11-15 |
|  | 21924864 | 40359 | 06 35 09.2 +03 41 20 | 0.007 | 31.2 | 2007-11-25 |
|  | 3960064 | 37 | 06 32 18.0 +04 52 00 | 0.29 | 20.8 | 2004-03-09 |
| W 3 | 05050624 | 127 | 02 26 34.4 +62 00 42 | 0.26 | 62.4 | 2004-01-10 |
| W 4 | 13846016 | 20052 | 02 32 42.0 +61 27 00 | 0.48 | 52.0 | 2006-09-20 |
| M 17 | 12488704 | 107 | 18 20 30.0 −16 10 10 | 0.22 | 31.2 | 2005-09-16 |
| NGC 1893 | 15850240 | 20818 | 05 22 50.0 +33 25 00 | 0.21 | 52.0 | 2006-03-25 |

Note. — Column 1: Target name. Column 2: Astronomical Object Request number. Column 3: *Spitzer* program identification number. Column 4: AOR central right ascension and declination for epoch (J2000.0). Column 5: The total area of the AOR field of view. Column 6: Total integration time per pixel for long frames. Column 7: Date of the start of the observation.

Table 2.  GLIMPSE W 3 IRAC Photometry

| Designation | Catalog | α (J2000.0) (deg) | δ (J2000.0) (deg) | [3.6] (mag) | [4.5] (mag) | [5.8] (mag) | [8.0] (mag) |
|---|---|---|---|---|---|---|---|
| (1) | (2) | (3) | (4) | (5) | (6) | (7) | (8) |
| G133.7806+00.9202 | C | 36.3271765 | +61.7986413 | 14.62±0.03 | 14.57±0.04 | 14.14±0.12 | · · · |
| G133.7807+01.1976 | C | 36.5370775 | +62.0578550 | 14.60±0.05 | 14.12±0.05 | · · · | · · · |
| G133.7807+00.8824 | C | 36.2989563 | +61.7633080 | 16.79±0.06 | · · · | · · · | · · · |
| G133.7807+01.2416 | C | 36.5707385 | +62.0989199 | 10.04±0.02 | 9.91±0.02 | 9.79±0.03 | 9.85±0.09 |
| G133.7809+01.3590 | C | 36.6612036 | +62.2084902 | · · · | 14.34±0.06 | · · · | · · · |
| G133.7809+01.1562 | C | 36.5059422 | +62.0191577 | 15.16±0.06 | 15.20±0.13 | · · · | · · · |
| G133.7809+01.2510 | C | 36.5783478 | +62.1076071 | 10.93±0.02 | 10.91±0.02 | 10.88±0.04 | · · · |
| G133.7809+00.7947 | C | 36.2339929 | +61.6811519 | 16.20±0.06 | · · · | · · · | · · · |
| G133.7810+01.2832 | C | 36.6031637 | +62.1376556 | 14.72±0.08 | 14.65±0.08 | · · · | · · · |
| G133.7810+01.1437 | C | 36.4966881 | +62.0074271 | 16.36±0.06 | 16.07±0.09 | · · · | · · · |

Note. — Column 1: GLIMPSE source designation. Column 2: A - source is in the GLIMPSE Archive, C - source is in the GLIMPSE Catalog. Columns 3-4: Right ascension and declination for epoch (J2000.0). Columns 5-8: GLIMPSE IRAC magnitudes.

Table 3.   IRAC  Aperture  Photometry

| Target | Designation | $\alpha$ (J2000.0) (deg) | $\delta$ (J2000.0) (deg) | [3.6] (mag) | [4.5] (mag) | [5.8] (mag) | [8.0] (mag) | Aperture (pixels) |
|---|---|---|---|---|---|---|---|---|
| (1) | (2) | (3) | (4) | (5) | (6) | (7) | (8) | (9) |
| Flame | G206.1095-16.5153 | 85.0980833 | -1.6195000 | 16.24±0.11 | 15.84± 0.12 | ⋯ | ⋯ | 2 |
| Flame | G206.7464-16.8365 | 85.0982500 | -2.3089444 | 16.78±0.14 | 15.48± 0.07 | ⋯ | ⋯ | 4 |
| Flame | G206.7375-16.8311 | 85.0990000 | -2.2989167 | 16.74±0.11 | 16.43± 0.13 | ⋯ | ⋯ | 4 |
| Flame | G206.6196-16.7713 | 85.0995417 | -2.1711111 | 17.27±0.17 | 16.58± 0.07 | ⋯ | ⋯ | $2^a$ |
| Flame | G205.9689-16.4401 | 85.1014583 | -1.4651111 | 12.83±0.02 | 12.84± 0.02 | 12.62±0.08 | 12.05±0.08 | 4 |
| Flame | G206.1113-16.5122 | 85.1016250 | -1.6195278 | 15.97±0.08 | 15.68± 0.10 | ⋯ | ⋯ | 4 |
| Flame | G206.8668-16.8928 | 85.1018750 | -2.4371389 | 16.24±0.08 | 16.26± 0.09 | ⋯ | ⋯ | 2 |
| Flame | G206.3476-16.6308 | 85.1025000 | -1.8751667 | 10.58±0.02 | 10.31± 0.02 | 9.85±0.02 | 8.98±0.02 | 4 |
| Flame | G206.7471-16.8317 | 85.1028333 | -2.3073056 | 16.57±0.11 | 16.11± 0.10 | ⋯ | ⋯ | 4 |
| Flame | G205.8411-16.3730 | 85.1033333 | -1.3253889 | 16.36±0.13 | 16.90± 0.20 | ⋯ | ⋯ | 4 |

Note. — Column 1: Target name. Column 2: Source Designation. Columns 3-4: Right ascension and declination for epoch (J2000.0). Columns 5-8: IRAC magnitudes from aperture photometry. Column 9: Aperture size in pixels used for photometric extraction. Table 3 is published in its entirety in an electronic form. A portion is shown here for guidance regarding its form and content.

[a] Contaminating flux from bright, neighboring sources is $\geqq$10% of the source flux in the 2 pixel aperture.

Table 4. IRAC Catalog Properties

| Target | Location (l, b) | Area deg$^2$ | [3.6] (mag) | [4.5] (mag) | [5.8] (mag) | [8.0] (mag) | Sources |
|--------|-----------------|--------------|-------------|-------------|-------------|-------------|---------|
| (1) | (2) | (3) | (4) | (5) | (6) | (7) | (8) |
| **Aperture Photometry Catalogs** | | | | | | | |
| Flame | 206.5−16.4 | 1.27 | 16.5 | 16.3 | 13.3 | 13.1 | 18,185 |
| W 40 | 28.8+03.5 | 2.37 | 16.3 | 16.3 | 14.5 | 14.1 | 475,903 |
| RCW 36 | 265.1+01.4 | 0.03 | 14.7 | 14.1 | 11.3 | 9.1 | 723 |
| NGC 2264 | 203.0+02.2 | 0.50 | 16.5 | 16.3 | 14.3 | 13.1 | 24,539 |
| NGC 2362 | 238.2−05.6 | 0.50 | 16.7 | 16.7 | 14.7 | 13.7 | 16,559 |
| DR 21 | 81.7+00.5 | 1.08 | 16.1 | 16.1 | 13.9 | 13.3 | 139,887 |
| Rosette | 206.3−02.1 | 2.08 | 16.5 | 16.3 | 14.3 | 13.1 | 31070 |
| W 4 | 134.7+00.9 | 0.48 | 16.7 | 16.9 | 14.9 | 13.5 | 38,540 |
| NGC 1893 | 173.6−01.7 | 0.15 | 17.1 | 17.1 | 14.5 | 13.3 | 12,460 |
| **GLIMPSE Pipeline Catalogs** | | | | | | | |
| Lagoon | 6.0−01.2 | 0.73 | 13.1 | 13.1 | 11.9 | 11.3 | 289,844 |
| RCW 38 | 268.0−01.0 | 0.97 | 13.9 | 13.7 | 12.3 | 11.9 | 16,019 |
| NGC 6334 | 351.1+00.7 | 0.73 | 13.7 | 13.5 | 11.7 | 11.1 | 331,442 |
| NGC 6357 | 353.0+00.9 | 0.73 | 13.5 | 13.3 | 11.7 | 11.1 | 385,896 |
| Eagle | 17.0+00.8 | 0.60 | 13.7 | 13.5 | 11.7 | 10.9 | 102,876 |
| M 17 | 15.1−00.7 | 1.11 | 13.5 | 13.3 | 11.7 | 10.9 | 333,864 |
| W 3 | 133.9+01.1 | 0.25 | 14.9 | 15.5 | 12.5 | 10.7 | 10,733 |
| Trifid | 7.0−00.3 | 0.36 | 13.3 | 13.3 | 11.7 | 10.9 | 95,038 |
| NGC 3576 | 291.3−00.7 | 1.00 | 13.9 | 13.7 | 12.1 | 11.9 | 45,879 |

Note. — Column 1: Target name. Column 2: Galactic coordinates. Column 3: Angular area of IRAC mosaic field of view. Columns 4-7: Completeness limits for catalog sources (bands [3.6], [4.5], [5.8], and [8.0], respectively). Column 8: The number of point sources in the catalog.

## Table 5.   Aperture Size

| Δmag | 1.0 pix | 2.0 pix | 3.0 pix | 4.0 pix | 5.0 pix | 6.0 pix | 8.0 pix | 10.0 pix | 12.0 pix | 14.0 pix | 16.0 pix | 18.0 pix | 20.0 pix |
|---|---|---|---|---|---|---|---|---|---|---|---|---|---|
| 7.0 mag | 4; 0% | 4; 0% | 4; 0% | 4;0% | 4; 0% | 4; 0% | 4; 0% | 4; 0% | 4; 0% | 4; 0% | 4; 0% | 4; 0% | 4; 0% |
| 6.5 mag | 4; 0% | 4; 0% | 4; 0% | 4; 0% | 4; 0% | 4; 0% | 4; 0% | 4; 0% | 4; 0% | 4; 0% | 4; 0% | 4; 0% | 4; 0% |
| 6.0 mag | 4; 0% | 4; 0% | 4; 0% | 4; 0% | 4; 0% | 4; 0% | 4; 0% | 4; 0% | 4; 0% | 4; 0% | 4; 0% | 4; 0% | 4; 0% |
| 5.5 mag | 4;1% | 4;1% | 4; 0% | 4; 0% | 4; 0% | 4; 0% | 4; 0% | 4; 0% | 4; 0% | 4; 0% | 4; 0% | 4; 0% | 4; 0% |
| 5.0 mag | 4;1% | 4;1% | 4;1% | 4; 0% | 4; 0% | 4; 0% | 4; 0% | 4; 0% | 4; 0% | 4; 0% | 4; 0% | 4; 0% | 4; 0% |
| 4.5 mag | 4;2% | 4;1% | 4;1% | 4; 0% | 4; 0% | 4; 0% | 4; 0% | 4; 0% | 4; 0% | 4; 0% | 4; 0% | 4; 0% | 4; 0% |
| 4.0 mag | 4;2% | 4;2% | 4;2% | 4;1% | 4; 0% | 4; 0% | 4; 0% | 4; 0% | 4; 0% | 4; 0% | 4; 0% | 4; 0% | 4; 0% |
| 3.5 mag | 4;4% | 4;3% | 4;3% | 4;1% | 4; 0% | 4; 0% | 4; 0% | 4; 0% | 4; 0% | 4; 0% | 4; 0% | 4; 0% | 4; 0% |
| 3.0 mag | 2;4% | 3;4% | 4;4% | 4;2% | 4;1% | 4; 0% | 4; 0% | 4; 0% | 4; 0% | 4; 0% | 4; 0% | 4; 0% | 4; 0% |
| 2.5 mag | 2;9% | 2;2% | 3;3% | 4;4% | 4;1% | 4; 0% | 4; 0% | 4; 0% | 4; 0% | 4; 0% | 4; 0% | 4; 0% | 4; 0% |
| 2.0 mag | 2;12% | 2;3% | 2;3% | 3;1% | 4;3% | 4; 0% | 4; 0% | 4; 0% | 4; 0% | 4; 0% | 4; 0% | 4; 0% | 4; 0% |
| 1.5 mag | 2;19% | 2;8% | 2;2% | 3;3% | 4;3% | 4;1% | 4; 0% | 4; 0% | 4; 0% | 4; 0% | 4; 0% | 4; 0% | 4; 0% |
| 1.0 mag | 2;30% | 2;14% | 2; 0% | 2;2% | 3;1% | 4;2% | 4; 0% | 4; 0% | 4; 0% | 4; 0% | 4; 0% | 4; 0% | 4; 0% |
| 0.5 mag | 2; 49% | 2; 24% | 2; 4% | 2; 3% | 3; 1% | 4; 3% | 4; 0% | 4; 0% | 4; 0% | 4; 0% | 4; 0% | 4; 0% | 4; 0% |
| 0.0 mag | 2; 77% | 2; 45% | 2; 1% | 3; 3% | 3; 0% | 4; 0% | 4; 0% | 4; 0% | 4; 0% | 4; 0% | 4; 0% | 4; 0% | 4; 0% |
| -0.5 mag | 2; >100% | 2; 67% | 2; 17% | 2; 0% | 2; 3% | 3; 1% | 4; 1% | 4; 0% | 4; 0% | 4; 0% | 4; 0% | 4; 0% | 4; 0% |
| -1.0 mag | 2; >100% | 2; >100% | 2; 55% | 2; 11% | 2; 3% | 3; 2% | 4; 2% | 4; 0% | 4; 0% | 4; 0% | 4; 0% | 4; 0% | 4; 0% |
| -1.5 mag | 2; >100% | 2; >100% | 2; 34% | 2; 8% | 2; 2% | 3; 1% | 4; 3% | 4; 1% | 4; 0% | 4; 0% | 4; 0% | 4; 0% | 4; 0% |
| -2.0 mag | 2; >100% | 2; >100% | 2; 98% | 2; 26% | 2; 1% | 2; 5% | 3; 1% | 4; 3% | 4; 1% | 4; 0% | 4; 0% | 4; 0% | 4; 0% |
| -2.5 mag | 2; >100% | 2; >100% | 2; >100% | 2; 36% | 2; 2% | 2; 6% | 3; 2% | 4; 3% | 4; 1% | 4; 1% | 4; 0% | 4; 0% | 4; 0% |
| -3.0 mag | 2; >100% | 2; >100% | 2; >100% | 2; 76% | 2; 15% | 2; 3% | 2; 2% | 3; 0% | 4; 3% | 4; 2% | 4; 0% | 4; 0% | 4; 0% |
| -3.5 mag | 2; >100% | 2; >100% | 2; >100% | 2; 95% | 2; 18% | 2; 1% | 2; 3% | 3; 4% | 3; 1% | 4; 1% | 4; 0% | 4; 0% | 4; 0% |
| -4.0 mag | 2; >100% | 2; >100% | 2; >100% | 2; >100% | 2; 23% | 2; 6% | 2; 2% | 2; 5% | 3; 0% | 4; 3% | 4; 2% | 4; 1% | 4; 0% |
| -4.5 mag | 2; >100% | 2; >100% | 2; >100% | 2; >100% | 2; 79% | 2; 20% | 2; 0% | 3; 3% | 3; 3% | 3; 0% | 4; 4% | 4; 4% | 4; 1% |
| -5.0 mag | 2; >100% | 2; >100% | 2; >100% | 2; >100% | 2; 88% | 2; 31% | 2; 11% | 2; 5% | 2; 3% | 2; 4% | 2; 1% | 4; 3% | 4; 3% |
| -5.5 mag | 2; >100% | 2; >100% | 2; >100% | 2; >100% | 2; >100% | 2; 74% | 2; 19% | 2; 5% | 2; 3% | 2; 4% | 2; 1% | 4; 3% | 4; 3% |
| -6.0 mag | 2; >100% | 2; >100% | 2; >100% | 2; >100% | 2; >100% | 2; >100% | 2; 28% | 2; 13% | 2; 0% | 2; 4% | 2; 2% | 3; 0% | 3; 0% |
| -6.5 mag | 2; >100% | 2; >100% | 2; >100% | 2; >100% | 2; >100% | 2; >100% | 2; 34% | 2; 17% | 2; 10% | 2; 4% | 2; 3% | 3; 4% | 3; 3% |
| -7.0 mag | 2; >100% | 2; >100% | 2; >100% | 2; >100% | 2; >100% | 2; >100% | 2; 48% | 2; 54% | 2; 27% | 2; 5% | 2; 4% | 2; 2% | 3; 4% |

Note. — For each pair of magnitude difference (first column) and angular separation (first row), the chosen aperture size (in pixels) and the simulated error in flux measurement for that aperture size (rounded to the nearest percentage) are given.

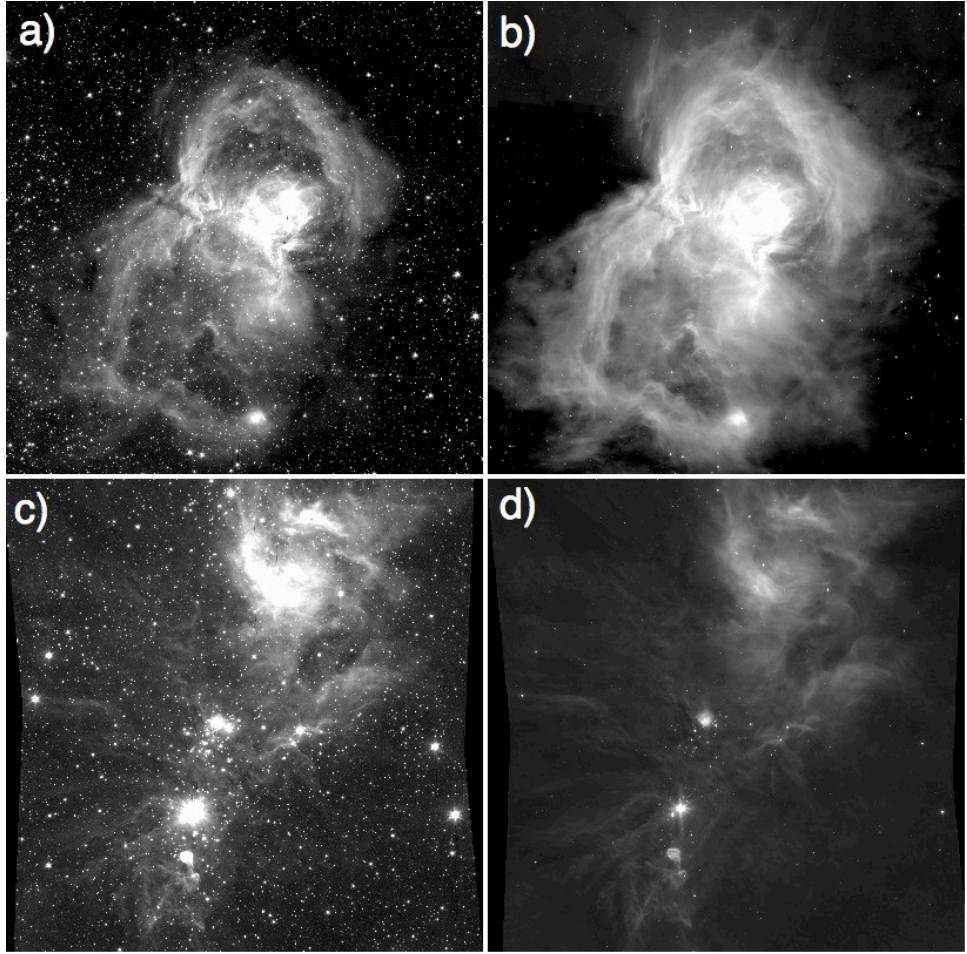

Fig. 1.— IRAC mosaics. (a) 3.6 $\mu$m band image of W 40. (b) 8.0 $\mu$m band image of W 40. (c) 3.6 $\mu$m band image of NGC 2264. (d) 8.0 $\mu$m band image of NGC 2264. Both regions have structurally complex nebulosity; however, nebulosity is higher in W 40 than in NGC 2264.

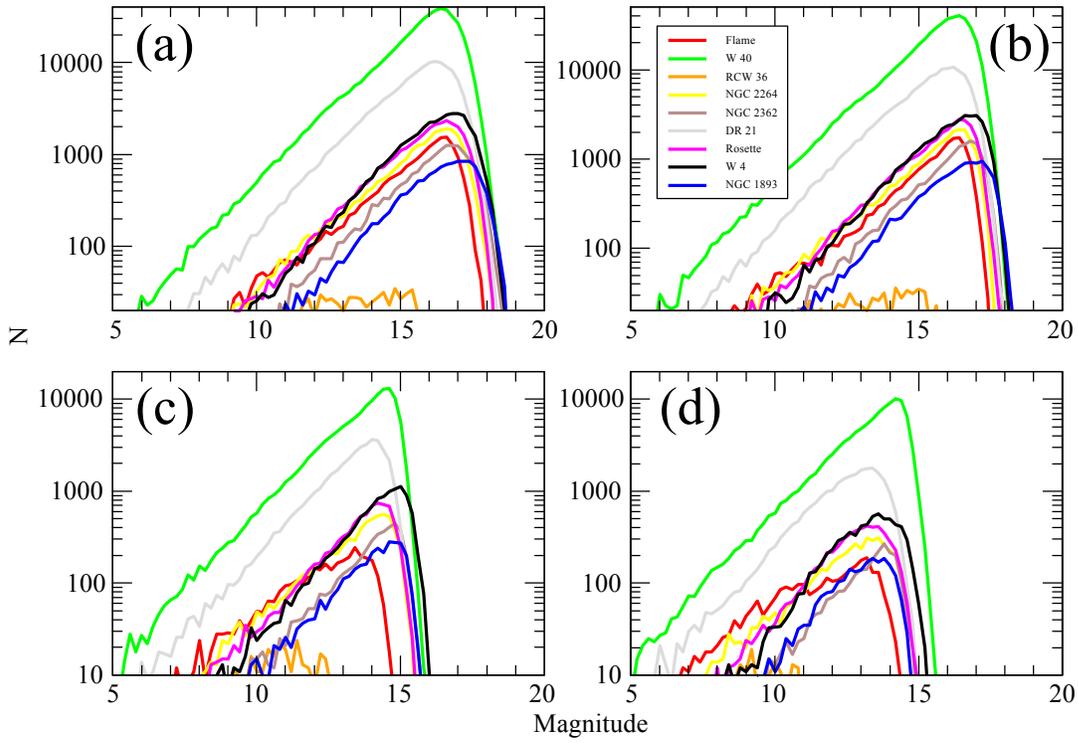

Fig. 2.— Histograms of 3.6 (a), 4.5 (b), 5.8 (c), and 8.0 $\mu$m (d) band point-source magnitudes for nine MYStIX regions using a bin width of 0.2 mag. The completeness limit is assumed to be one bin brighter than the peak of the histogram.

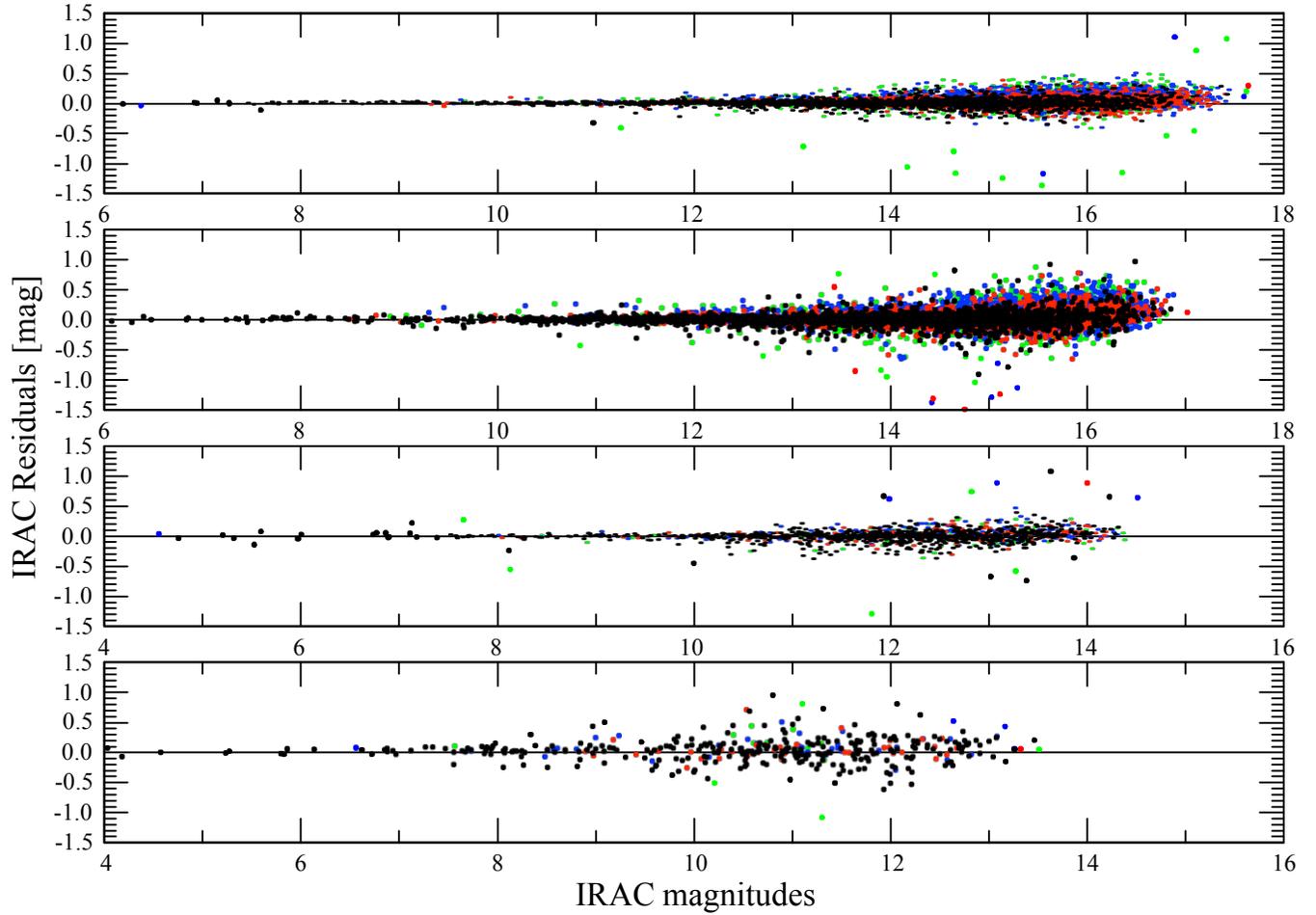

Fig. 3.— A comparison of aperture photometry to GLIMPSE photometry for W 3. Plot of magnitude residuals (aperture minus GLIMPSE) vs. magnitude using aperture and GLIMPSE catalogs produced from the HDR *Spitzer* observation of the W 3 region. Plots (*top* to *bottom*) are the 3.6, 4.5, 5.8, and 8.0 $\mu$m bands, respectively. Colors indicate the aperture used: black for the 4 pixel aperture, red for the 3 pixel aperture, blue for the 2 pixel aperture with low crowding, and green for the 2 pixel aperture with high crowding.

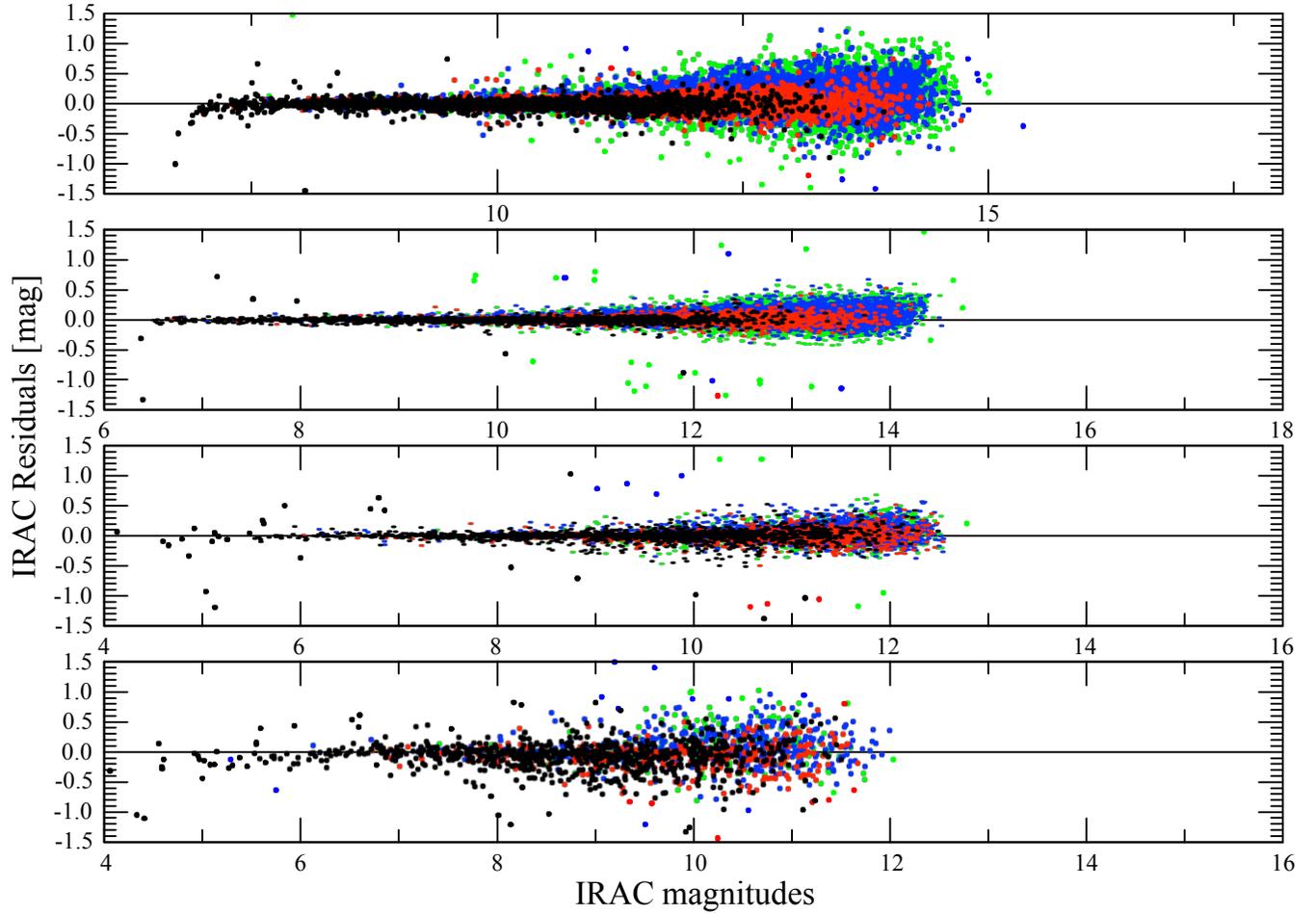

Fig. 4.— A comparison of aperture photometry to GLIMPSE photometry for M 17, similar to Figure 3. Plots (*top* to *bottom*) are the 3.6, 4.5, 5.8, and 8.0 $\mu$m bands, respectively. Here, the M 17 GLIMPSE data comes from significantly shorter observations than the aperture-photometry data, leading to more scatter than is seen in the comparison for W 3.

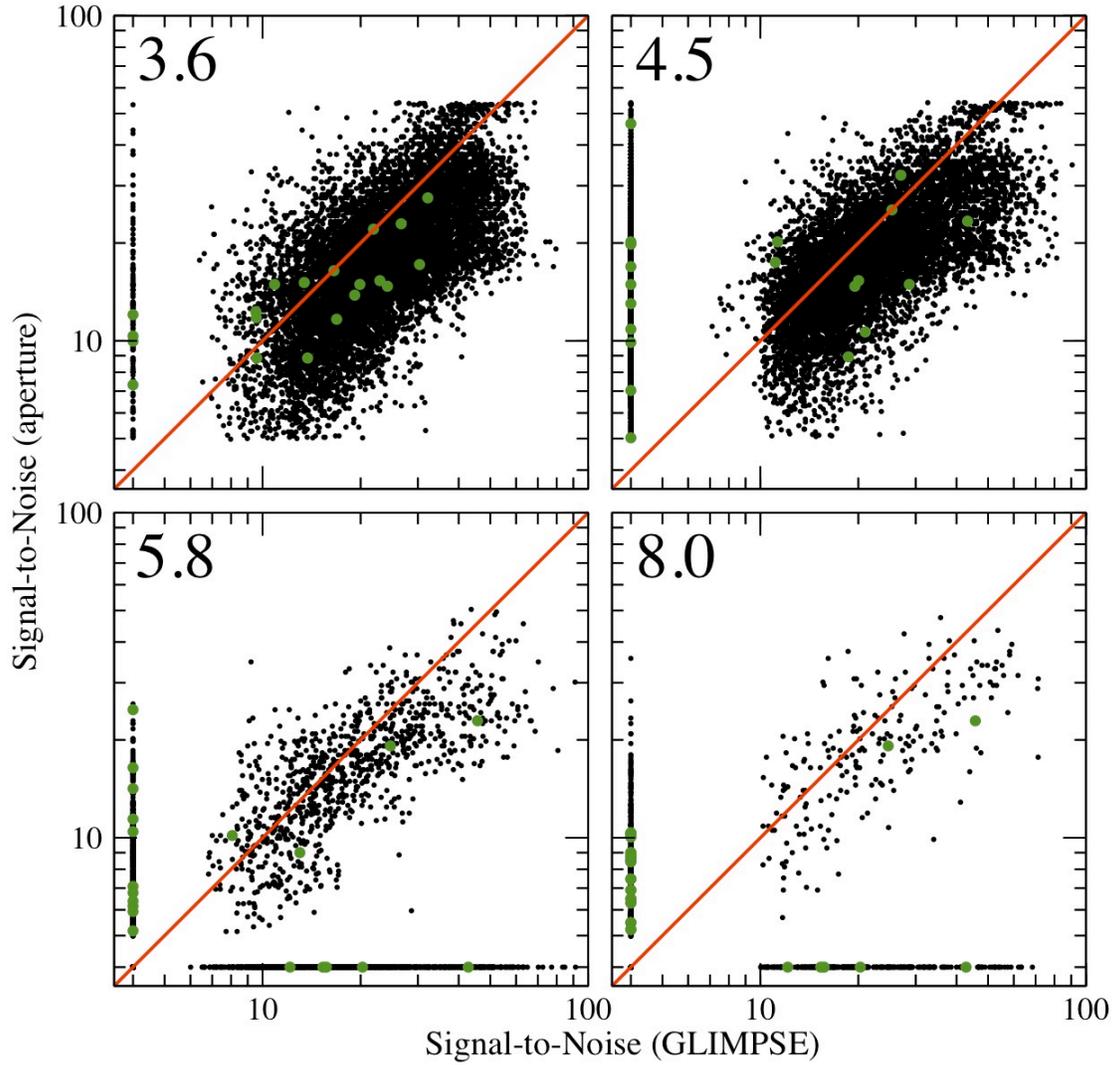

Fig. 5.— Signal-to-noise of the aperture photometry catalog vs. signal-to-noise in the GLIMPSE catalog for all W 3 sources present in both catalogs for each band; non-detections are set to S/N=4 on the graph. The green points are the sources flagged as possible spurious detections of nebulosity. The red line indicates equal error from both catalogs.

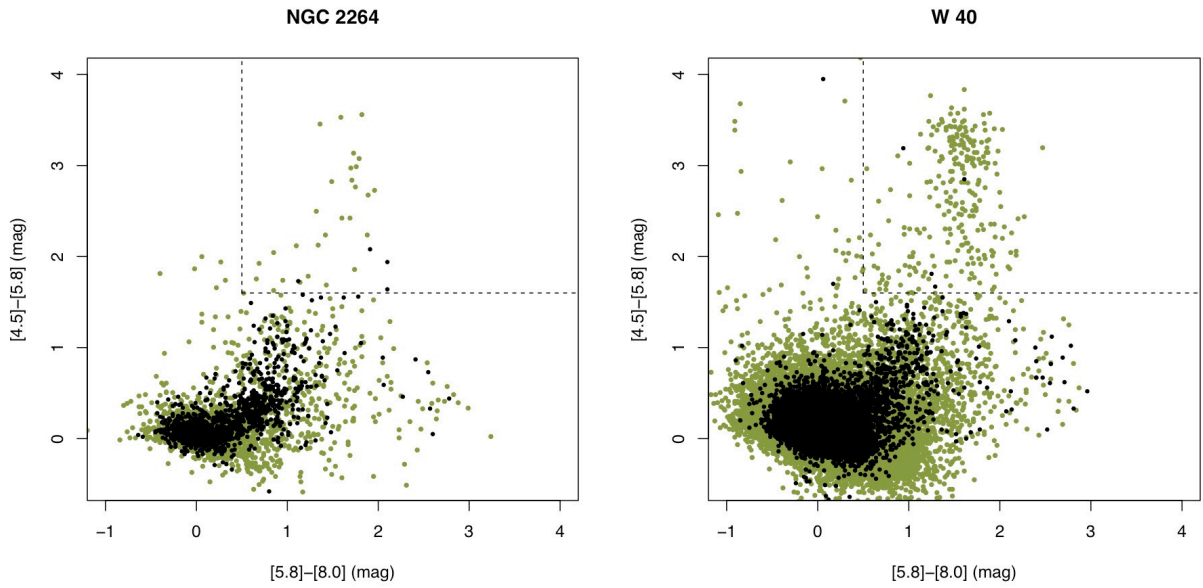

Fig. 6.— [4.5]-[5.8] vs. [5.8]-[8.0] color-color diagrams − *Left:* NGC 2264, *Right:* W 40. Sources with *S/N* > 10 in all four IRAC bands are black circles and sources with *S/N* < 10 in at least one band are green circles. The dashed lines show the color cuts for PAH knots from Povich et al. (2013).

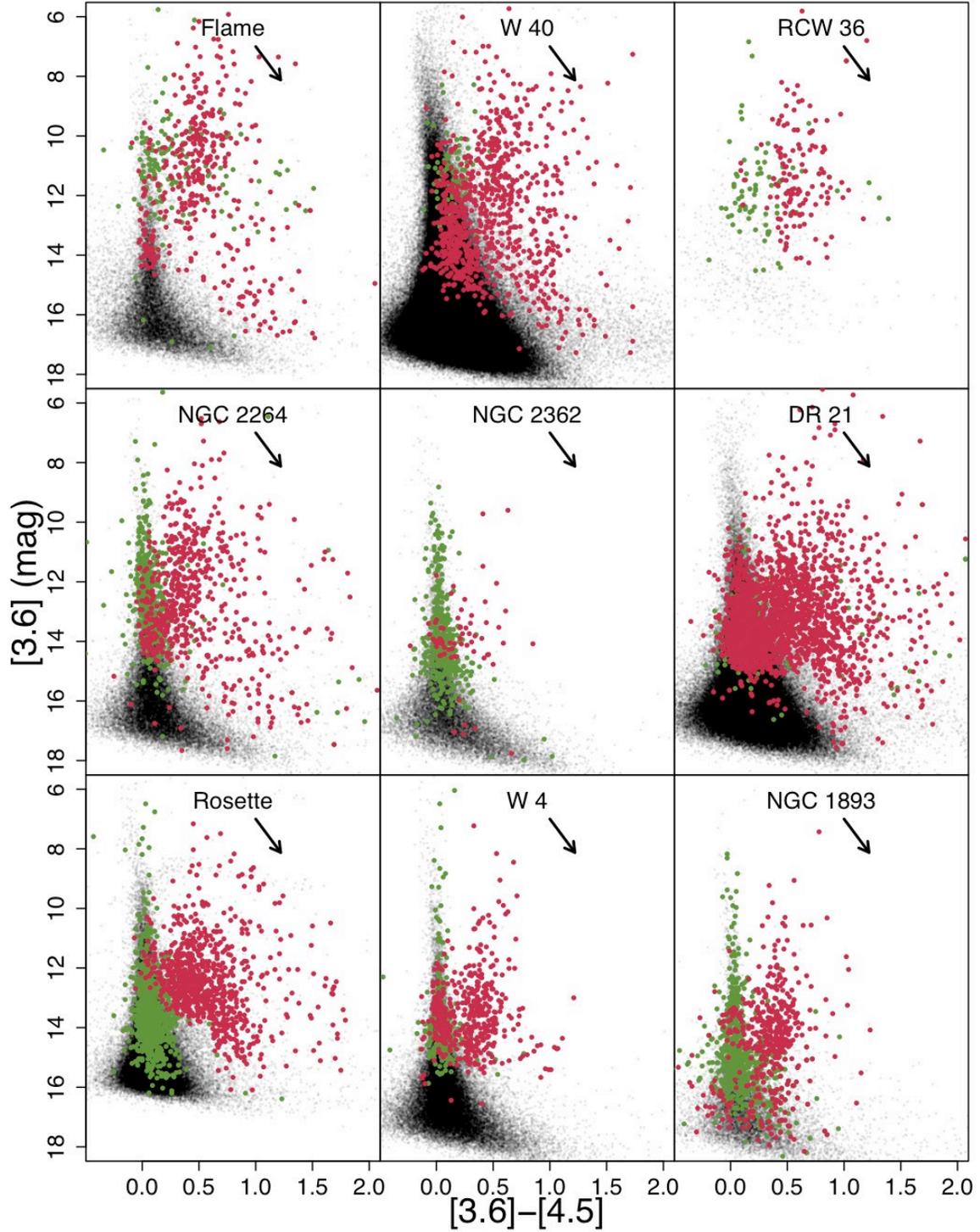

Fig. 8.— [3.6] vs. [3.6]-[4.6] color-magnitude diagrams for MIR sources in the each region. MPCMs with no IRE are shown by green circles, MPCMs with IRE are shown by red circles, and remaining sources (field stars, unclassified members, and non-stellar sources) are shown by grey circles. The arrow indicates reddening of $A_K = 2$ mag using the Indebetouw et al. (2005) extinction law.

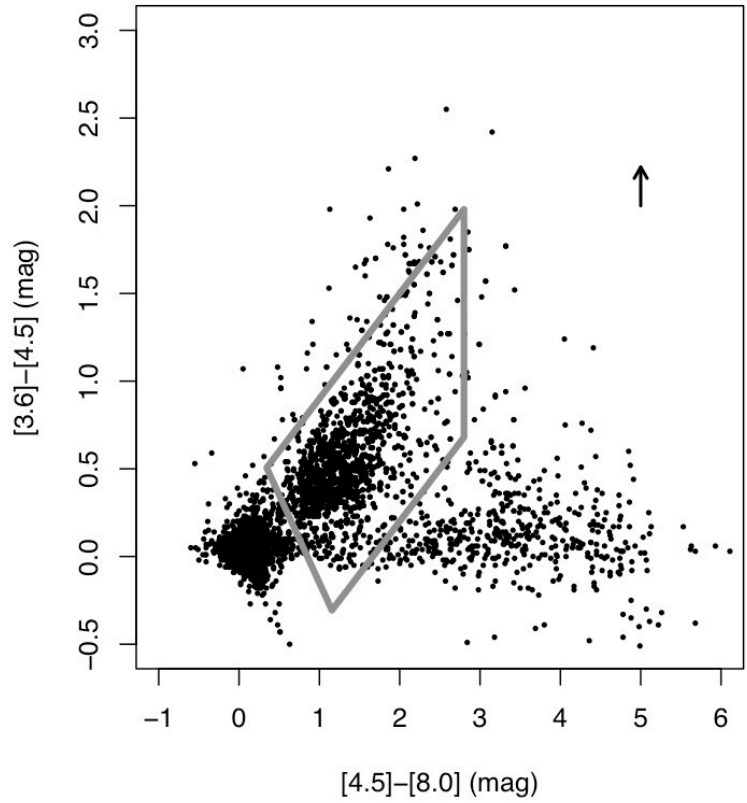

Fig. 9.— $[3.6] - [4.5]$ vs. $[4.5] - [8.0]$ color-color diagrams for MIR sources in the Rosette Nebula. The color cuts used by Simon et al. (2007) are shown with the gray lines. These cuts identify a slightly different set of IRE sources compared to Povich et al. (2013). The black arrow is the $A_K = 2$ mag reddening vector.